\documentclass[final,5p,times,twocolumn]{elsarticle}
\usepackage{amsmath}
\usepackage{amssymb}
\usepackage{lipsum}
\usepackage{siunitx}

\DeclareSIUnit\atomicmassunit{u}

\journal{Nuclear Physics A}

\begin{document}

\begin{frontmatter}



% \title{First Application of the Deep Lie Map Network Method to Identify Magnetic Field Errors in a Synchrotron}
\title{Beam-based Identification of Magnetic Field Errors in a Synchrotron using Deep Lie Map Networks}

\author[first]{Conrad Caliari}
\author[second]{Adrian Oeftiger}
\author[first,second]{Oliver Boine-Frankenheim}

\affiliation[first]{
            % organization={TEMF, Technische Universitat Darmstadt}
            organization={Institute for Accelerator Science and Electromagnetic Fields (TEMF), Technische Universität Darmstadt},%Department and Organization
            addressline={Schlossgartenstraße 8}, 
            city={Darmstadt},
            postcode={64289}, 
            state={},
            country={Germany}
}

\affiliation[second]{organization={GSI Helmholtzzentrum für Schwerionenforschung GmbH},%Department and Organization
            addressline={Planckstraße 1}, 
            city={Darmstadt},
            postcode={64291}, 
            state={},
            country={Germany}}

\begin{abstract}
We present the first experimental validation of the Deep Lie Map Network (DLMN) approach for recovering both linear and non-linear optics in a synchrotron.
The DLMN facilitates the construction of a detailed accelerator model by integrating charged particle dynamics with machine learning methodology in a data-driven framework.
The primary observable is the centroid motion over a limited number of turns, captured by beam position monitors.
The DLMN produces an updated description of the accelerator in terms of magnetic multipole components, which can be directly utilized in established accelerator physics tools and tracking codes for further analysis.
In this study, we apply the DLMN to the SIS18 hadron synchrotron at GSI for the first time.

We discuss the validity of the recovered linear and non-linear optics, including quadrupole and sextupole errors, and compare our results with alternative methods, such as the LOCO fit of a measured orbit response matrix and the evaluation of resonance driving terms.
The small number of required trajectory measurements, one for linear and three for non-linear optics reconstruction, demonstrates the method's time efficiency.
Our findings indicate that the DLMN is well-suited for identifying linear optics, and the recovery of non-linear optics is achievable within the capabilities of the current beam position monitor system.
We demonstrate the application of DLMN results through simulated resonance diagrams in tune space and their comparison with measurements.
The DLMN provides a novel tool for analyzing the causal origins of resonances and exploring potential compensation schemes.
\end{abstract}

%\includegraphics[width=\columnwidth]{grabs}



\end{frontmatter}

%\tableofcontents



\section{Introduction}
\label{Section:Introduction}
The ever growing demand for high intensity and high brilliance hadron beams by various users ranging from nuclear physics to anti-matter studies has increased synchrotron performance requirements ever since. 
Magnetic field errors originating from feedown caused by misalignments, fabrication errors or power supply failures are detrimental to beam quality, as resonances may be excited or instability thresholds crossed resulting in beam loss. In order to operate synchrotrons close to their performance limitations, an accurate accelerator model including both linear and non-linear magnetic field errors is desirable.
The detailed study of the interplay of magnetic resonances, motion in 6D phase space and collective effects like space charge forces necessitate numerical simulation studies and, thus, detailed computer models.
Realistic computer models can be developed through methods that attribute the discrepancies between the nominal performance of an accelerator and experimental observations to underlying physical quantities and machine properties, such as magnetic multipole components.
This is in particular useful to operate hadron synchrotrons where space-charge induced resonance crossing is a major intensity limitation like the SIS100 \cite{Oeftiger:SpaceChargeLimit}. The SIS100 is going to be the central workhorse of the FAIR project aiming at highest intensities of heavy-ion beams \cite{Spiller:FAIR}.  

The SIS18 synchrotron, first commissioned in 1990, is going to serve as an injector to SIS100 \cite{Ondreka:Recommissioning}. A systematic shift in betatron tune and a discrepancy in chromaticity between the nominal optics model serving the control system and measurements is observed in operation. Furthermore, the presence of unexpected skew sextupole resonances is revealed by tune scans. Thus, SIS18 serves as a good candidate to improve the linear and non-linear optics model based on measurement data. It provides an interesting test bed to compare the newly developed approach of Deep Lie Map Networks (DLMN) to other, commonly employed optics measurement methods.

An established method to obtain an accelerator model including linear field errors is the linear optics from closed orbits (LOCO) method \cite{Safranek:LOCO}. LOCO requires the measurement of an orbit response matrix (ORM), which in linear approximation links the effect of steering magnets to the change in orbit. This measured ORM is fitted by a computer model, which yields both normal and skew quadrupole errors capable of explaining the measurement. Successful applications are reported for instance in \cite{Dowd:LOCO_at_ALS, Nadolski:LOCO_at_SOLEIL}.

Different methods to generate a model of non-linear magnetic field errors have been proposed. The measurement of resonance driving terms (RDT) estimates the strength of sextupole errors from the spectra of centroid oscillations induced by a kicker magnet. This method allows a localization of the error up to its neighboring beam position monitors \cite{Tomas:LocalResonanceTerms}, but suffers from decoherence due to chromatic and amplitude detuning.
The non-linear tune response (NTRM) method is based upon measurement of the change in tune caused by deforming the orbit globally \cite{Parfenova:NTRM}. Non-linear field errors yield a quadrupole-like error contribution via feed-down and, thus, change the betatron tune. The measured tune shift in dependence of steering magnets yields a matrix comparable to the ORM, which is then fitted by a computer model. This method is limited by the accuracy of tune measurements and by its perturbative modelling approach which neglects effects like amplitude detuning.
Both the application of the RDT as well as the NTRM method for the measurement of non-linear optics require an accurate linear optics model.

The rapidly growing field of machine learning and artificial intelligence facilitates the search for new methods to generate accurate accelerator models from limited amount of data and, thus, in a beam time efficient manner.
In \cite{Ivanov:PNN_beamDynamics} a physics-inspired neural network (PNN) using Taylor Map (TM) layers, henceforth referred to as TM-PNN, is proposed to model synchrotron optics. The work successfully applies a TM-PNN to closed orbit correction. 
A generalization to linear and, further, non-linear particle accelerator optics is challenging due to the exponential growth of degrees of freedom.
Consequently, we observe over-fitting issues which could not be resolved by the proposed soft constraint of symplectic regularization \cite{Caliari:Master}.

In \cite{Caliari:DLMN} the Deep Lie Map Network (DLMN) approach is demonstrated in simulations, which applies training algorithms from the field of machine learning with conventional models from particle accelerator optics in order to improve an optics model in a data-driven way.
The DLMN yields both linear and non-linear magnetic field errors from only a few synchrotron shots.
It fits a computer model to the trajectory of the beam centroid, observed for few turns only by means of beam position monitors, after the beam is excited by a kicker magnet.
In contrast to TM-PNN, particle optics are modelled in drift-kick approximation which significantly reduces the number of degrees of freedom.
The approach has the potential to be very efficient in beam time as no time-consuming installation of orbit bumps along the accelerator is required. In contrast to RDT or NTRM measurements, which require data acquisition for few thousand turns in order to recover tunes and higher harmonics from spectra, the short time interval required by the DLMN measurement in the time domain leaves the method relatively unaffected by decoherence effects, as detailed in \cite{Caliari:DLMN}.

Time-efficient methods for modeling synchrotron optics increase an accelerator's uptime and may pave the way toward monitoring machine optics during regular operation.
This is of particular interest to synchrotrons that accelerate various ion species under rapidly changing user demands and machine settings.

The present work reports on the first application of the DLMN method to measurement data.
First results are demonstrated using a heavy-ion beam at the SIS18 booster synchrotron.
A linear optics model is recovered from a single trajectory only, and its plausibility gauged by comparison a spectral analysis of turn-by-turn BPM data and a LOCO fit of a measured orbit response matrix.
Non-linear optics are recovered from a dataset comprising three trajectories, and the DLMN's predictions are compared to an evaluation of resonance driving terms and to amplitude detuning observed in measurements.
We discuss the potential of this novel approach in comparison to established alternative approaches.

The remainder of this paper is structured as follows: Sec.~\ref{Section:DLMN} reviews the DLMN approach, Sec.~\ref{Section:SetupSIS18} describes the beam experiment at SIS18, the results are discussed in Sec.~\ref{Section:Results}, followed by a Conclusion.

\section{Deep Lie Map Networks}
\label{Section:DLMN}
The Deep Lie Map Network (DLMN) approach models the accelerator in the framework of Hamiltonian dynamics \cite{Caliari:DLMN}. The equations of motion can be approximately solved for sections with a magnetic field invariant along the reference path, i.e.\ beam line element by element, using the drift-kick approximation. The result is a concatenation of analytic maps, updating particle position and momentum in an alternating scheme. This modelling approach allows consideration of multipole components of arbitrary order. The only approximation made is the chosen split scheme into discrete coordinate updates, and split schemes exceeding arbitrary expansion orders are known \cite{Yoshida:SymplecticIntegrators}. Magnetic fringe fields are taken into account up to first order necessary to model the edge focusing of bending magnets. The motion of the beam is modeled by a single particle. Hence, decoherence of the centroid motion is not taken into account here. The DLMN is used to predict the centroid trajectory only for few turns, which is short compared to the synchrotron revolution period and, thus, rf-cavities can be neglected.

The drifts and kicks are given by analytic expressions, and can be differentiated analytically. Although possible, it is infeasible to apply the chain rule from calculus to the complex concatenation of many drifts and kicks because of expression swelling, which even overstrains computer codes capable of symbolic differentiation. The technique of automatic differentiation is well suited as it evaluates the chain rule by consecutively evaluating each coordinate update's derivative. The predicted trajectory can be differentiated w.r.t.\ the individual multipole components describing the magnetic field in each beam line element. The complexity of beam dynamics arises from the concatenation of many of these simple building blocks, an analogous situation to the structure of artificial neural networks. An artificial neural network maps some input vector $x$ to its output $y = \text{ANN}(x)$. Training of an ANN alters the networks weights such that the network prediction ($x$, $\text{ANN}(x)$) matches the observation ($x$, $y$), on average and over many observations. In case of the DLMN, the magnetic multipole components are understood as weights, the input $x$ is a phase space vector serving as initial condition and the observation $y$ is a centroid trajectory measured by means of beam position monitors.

This analogy allows implementation of the DLMN model in common machine learning frameworks and utilization of established training algorithms. The training results presented in this work are obtained by means of the ADAM algorithm \cite{Adam}, a gradient-descent based optimization algorithm that scales parameter updates based on the running average of past gradients. 

The DLMN is initialized by means of the nominal accelerator optics model, which originates from known multipole strengths of the magnets and their location in the tunnel. 
 It defines the desired accelerator status and underlies the accelerators control system.
Unknown discrepancies like misalignments, fabrication errors, ground motion, stray fields from magnetic septa, power supply calibration errors, etc.\ result in a different accelerator status, perturbed particle optics and changed beam properties.
The DLMN method yields an effective machine model in terms of magnetic multipole components, taking into account the effect of these discrepancies as sensed by the beam.
It is therefore desirable to update the nominal accelerator model using beam-based optics measurements. This refined model may then be applied to shed light on the origin of those discrepancies in simulation studies and guide countermeasures like powering corrector magnets.

\subsection{Comparison to Alternative Methods}
Central to optics measurements are beam position monitors (BPMs) installed periodically in a synchrotron, which measure the transverse position of the beam centroid. Transverse beam profiles are more difficult to measure, and in case of SIS18 only a single intensity profile monitor (IPM) exists with limited spatial resolution \cite{Giacomini:IPM}. The only observable of transverse beam dynamics taken into account here is, therefore, the transverse beam position at a few discrete positions along the circumference where a BPM is installed.
A BPM measurement yields the beam centroid position as a time series.
In first order approximation of single particle dynamics, the centroid position $\langle z \rangle$ undergoes a sinusoidal oscillation around some fixed point, the closed orbit, with the tune $Q_z$ denoting its frequency. If not stated otherwise, $z$ represents either the horizontal $x$ or vertical plane $y$ in this manuscript.

Possible approaches to optics measurements are
\begin{enumerate}
    \item measuring the linearized change in orbit induced by dipole correctors $\rightarrow$ yields an orbit response matrix (ORM) that may be fitted by a computer model (LOCO)
    \item observing the linearized change in tune induced by orbit bumps $\rightarrow$ non-linear tune response matrix (NTRM)
    \item performing a spectral analysis of centroid motion to estimate amplitudes of harmonics of betatron motion $\rightarrow$ resonance driving terms (RDT)
    \item analysing the centroid motion along the circumference $\rightarrow$ TM-PNN and DLMN
\end{enumerate}

Each approach yields a different quantity which may be fitted by a computer model. Approaches 1) and 2) require the time-consuming installation of orbit bumps along the circumference. In contrast, approaches 3) and 4) observe the centroid motion without actively sweeping optics settings, but require a few different excitation strengths from the kicker to sample transverse amplitude. For the DLMN, we additionally vary an rf-frequency offset to sample a chromatic tune shift yielding a total of ten different settings.
Neglecting considerations w.r.t.\ to noise, this amounts to ten cycles required by the synchrotron. Therefore, these approaches may in principle be more efficient in terms of beam time. 

It is common to all approaches that they neglect effects originating from the finite phase space volume occupied by the beam distribution, such as non-linear amplitude and chromatic detuning, as well as collective effects. Such detuning effects result in damping of the centroid oscillation, commonly known as \emph{decoherence}.

Approach 1) is less affected by turn-by-turn BPM noise as it averages out over long sampling lengths applicable to closed orbit measurements.
In contrast, this approach is subject to shot-by-shot fluctuations of the closed orbit.
In case of SIS18, the closed orbit rms orbit fluctuation across different cycles amounts to $\approx$\SI{100}{\micro\meter}, not to be confused with the BPM turn-by-turn measurement uncertainty.
This approach is unique in that decoherence has no effect at all while, at the same, it is limited to recovering a linear optics model only in its original formulation LOCO. A generalization to second-order optics is possible by repeated measured of orbit response matrices at different beam energies called NOECO \cite{Olsson:NOECO}. The orbit response's energy dependence can be studied using perturbation theory to reveal sextupole errors.

Approach 2) relies on inference of the betatron tune from BPM turn-by-turn data with high precision. Therefore, advanced algorithms for frequency analysis such as NAFF \cite{Laskar:NAFF} or SUSSIX \cite{Bartolini:SUSSIX} are applied which reach a frequency resolution of the order of $1/N^3$ on non-noisy oscillation signals and thus, outperform the frequency resolution of a Fast-Fourier Transform (FFT).
While obeying a similar scaling law for the resolution w.r.t.\ to the number of turns, further improvement in tune resolution is achievable by the analyzing the beam motion using many beam position monitors \cite{Zisopoulos:MixedBPM_NAFF}.

If the BPM readings are subject to significant noise, the frequency resolution of the advanced algorithms can be significantly reduced down toward the one achievable by an FFT \cite{Biancacci:FFTCorrectionTune}, which is of order $1/N$ for a signal of sample length $N$.
The applicability of this approach is, thus, restricted by two factors: decoherence, which limits the sample length $N$, and the signal-to-noise ratio of the BPM system.

Approach 3) performs a spectral analysis of centroid oscillations and is subject to turn-by-turn BPM noise. Assuming the noise to be of white Gaussian type $\propto \mathcal{N}(0,\sigma)$, the amplitude of each spectral line follows a Rayleigh distribution \cite{Bendat:NoiseDistribution}
\begin{align}
    |a_k| \propto \text{Rayleigh} \left( s = \frac{\sigma}{\sqrt{2 N}} \right) \; ,
    \label{Eqn:RDTNoiseFloorDistribution}
\end{align}
where the scale parameter $s$ depends on the standard deviation of the white noise and the sample length of $N$ turns. In order to measure resonance driving terms, the corresponding spectral line must be distinguishable from background noise floor given by Eq.~\eqref{Eqn:RDTNoiseFloorDistribution}. In contrast, to lower the noise floor by sampling for many turns, the number of turns $N$ is limited by the decoherence length. Furthermore, decoherence prohibits determination of the actual amplitudes, complicating a fit by a computer model.
Decoherence can be suppressed by driving centroid oscillations using an ac-dipole.
A switch from free to driven betatron oscillations introduces systematic changes to the driving terms, complicating the theoretical treatment and interpretation of measurement results \cite{Tomas:ACDipole}.

Approaches 4) aim at matching the time series obtained from the BPMs and thus are subject to BPM noise.
In contrast to approach 3), which performs a spectral analysis of the centroid motion, these methods do not benefit from the $1/\sqrt{2N}$ reduction in noise.
Conversely, the limitation to short time series comprising few turns only makes these approaches rather unaffected by decoherence and thus, allows to recover absolute values for betatron amplitudes.
A unique feature to approach 4) is the simultaneous reconstruction of linear and non-linear optics. Hence, the recovery of non-linearities does not require an \emph{a priori} established, accurate linear optics model. 

The DLMN approach in particular yields an optics model in terms of magnetic multipole components. This warrants a physical interpretation of the results, as well as further use by means of established tools and simulation codes in the accelerator physics community.

\section{Experiment at SIS18 Synchrotron}
\label{Section:SetupSIS18}
The DLMN method is trained on trajectories measured at the SIS18 synchrotron. Training the DLMN varies quadrupole and sextupole strengths in order to minimize the discrepancy between predicted and observed beam centroid motion. This discrepancy is quantified by the mean absolute error (MAE)
\begin{align}
    \text{MAE} = \frac{1}{L M N} \sum_{l=1}^L \sum_{m=1}^M \sum_{n=1}^N \left|\left| \left( \begin{bmatrix} x \\ y \end{bmatrix} - \begin{bmatrix} \hat{x} \\ \hat{y} \end{bmatrix} \right)_{l,m,n} \right|\right|_1 \quad , 
    \label{Eqn:MAE}
\end{align}
which compares the centroid position at $N$ BPMs over $M$ turns for $L$ different initial conditions. Here, $x$, $y$ denote the predicted horizontal and vertical centroid position, variables assigned a hat refer to measurements.

A successful training yields a DLMN capable of reproducing the measured motion of the beam centroid. A comparison of the magnetic field strengths between the nominal accelerator model and the more realistic, trained DLMN yields an error model of transverse optics.

Previous measurements at SIS18 revealed a systematic discrepancy in betatron tunes and chromaticity in both planes in comparison to the existing nominal accelerator model.
The SIS18 serves as a useful test case for applying the DLMN, as both linear and non-linear optics require improved modelling.

Training of the DLMN improves the quality of its predicted trajectories in that the mean absolute error (MAE) is continuously reduced. The validity of the optics model obtained from DLMN training is discussed in Section~\ref{Section:OpticsResults}.

\subsection{The SIS18 Synchrotron}
\begin{figure}
	\centering 
	\includegraphics[width=\columnwidth]{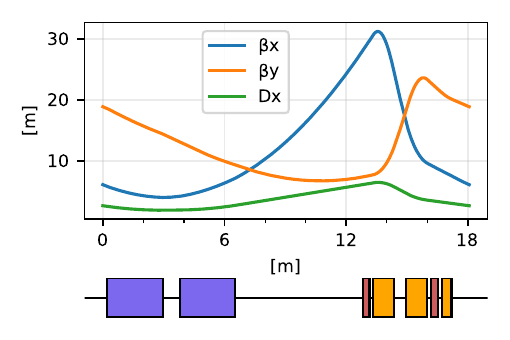}	
	\caption{A \SI{18}{\meter} long cell of SIS18 drawn to scale. Bending magnets are shown in violet, followed by the \textsf{KS1C} and \textsf{KS3C} sextupoles in red. The sextupoles enclose the orange quadrupoles \textsf{QS1F} and \textsf{QS2D}, the \textsf{QS3T} quadrupole is located at the cell end. The $\beta$-functions and horizontal dispersion are shown for the doublet-like optics.} 
	\label{Fig:SIS18_Cell_Layout}
\end{figure}

The DLMN model is applied to the heavy ion synchrotron SIS18 at GSI, Darmstadt, Germany. This \SI{216}{\meter} long accelerator ring is designed to accelerate ions ranging from protons to uranium, up to a maximum magnetic rigidity of \SI{18.5}{\tesla\meter}. SIS18 features twelve lattice cells which symmetrically repeat a configuration of two bending magnets and three quadrupoles per cell. The quadrupoles are grouped into a family of defocusing quadrupoles named \textsf{QS2D}, which is surrounded by two families of focusing quadrupoles, \textsf{QS1F} and \textsf{QS3T}. The SIS18 uses a triplet optic during injection to achieve a large admittance \cite{Franczak:SIS18}. During acceleration, the focusing strength of the \textsf{QS3T} family is reduced to about 10\% of the \textsf{QS1F} family, approaching a doublet optic. The reported experiments were all conducted at flattop using the doublet-like optics. Additionally, the six odd-numbered cells each comprise two sextupoles featuring individual power supplies. For chromaticity correction, the sextupoles are grouped into two families of focusing and defocusing sextupoles, referred to as  \textsf{KS1C} and \textsf{KS3C}, respectively. The layout of such an odd numbered cell together with nominal $\beta$-functions and horizontal dispersion function are displayed in Fig.~\ref{Fig:SIS18_Cell_Layout}. 

\subsection{Beam Properties}
Although the DLMN is trained on trajectories comprising a few turns only, decoherence of the centroid oscillation affects the resolution of magnetic field errors. The DLMN predicts trajectories of the beam centroid by modelling it by a single particle and hence, does not take into account decoherence due to chromatic or amplitude detuning. Previous simulation studies \cite{Caliari:DLMN} have shown the benefit of utilizing a ``pencil-like'' beam for measurements, i.e.\ a beam with small momentum spread and transverse emittances.

We report about our measurements performed with $\text{Au}^{65+}$ ions at an intermediate magnetic rigidity of \SI{5.5}{\tesla\meter}, which is a trade-off between adiabatically shrinking of the transverse emittances towards high energy and achieving a reasonably large amplitude from the excitation kick. The momentum spread is estimated from the longitudinal Hamiltonian $\mathcal{H}$ in small amplitude approximation. 

Sufficiently long after the ramp, the longitudinal phase space distribution becomes stationary.
This is supported by the noise analysis of transverse centroid position data, which does not reveal any residual centroid oscillations before excitation, c.f. Fig.~\ref{Fig:BPM_PreNoise_Procedure}.
The bunch length is measured by means of a fast current transformer (FCT) \cite{Chorniy:FCT_DAQ} and is linked to the rms-momentum spread assuming small amplitude synchrotron oscillations \cite{Lee:AcceleratorPhysics}.
We estimate the rms-momentum spread of the $\text{Au}^{65+}$ beam to $\sigma_\delta=\num{4.8e-4}$.

The non-normalized transverse emittances $\epsilon_x$, $\epsilon_y$ are inferred from the measured transverse beam size
\begin{align}
    \left\langle z^2\right\rangle = \sqrt{\beta_z \epsilon_z + D_z^2 \sigma_\delta^2} \quad,
\end{align}
where $\beta_z$, $D_z$ denote $\beta$-function and dispersion respectively, and $z$ denotes either the horizontal $x$ or vertical $y$ plane. The rms-beam size $\langle z^2 \rangle$ is obtained by measuring the transverse profile by means of an ionization profile monitor (IPM). The measured beam profile is approximately Gaussian and the normalized transverse emittances are estimated to $\epsilon_{x,\text{norm}}^\text{4-rms}=\SI{20}{\micro\meter}$ and $\epsilon_{y,\text{norm}}^\text{4-rms}=\SI{4}{\micro\meter}$.

An overview of SIS18 key properties is given in Table~\ref{tab:BeamParameters}.

\begin{table}[tb]
    \centering
    \caption{Properties of SIS18 and Key Beam Parameters.}

    \begin{tabular*}{\linewidth}{l@{\extracolsep{\fill}}c}
        \hline
        \hline
        Parameter & Value \\ \hline
        Circumference   &   \SI{216}{\meter}    \\
        Momentum compaction $\alpha_C$  &   \num{3.21e-2}    \\
        % Transition energy $\gamma_T$  &   5.5    \\
        Synchrotron tune $Q_s$   &   $\approx$ \num{1000} turns \\
        Set betatron tunes $Q_x$, $Q_y$  &   4.29, 3.29   \\
        Natural (absolute) chromaticity $\xi_x^\text{(nat)}$, $\xi_y^\text{(nat)}$    &   -6.43 / -4.89   \\
        Magnetic rigidity $(B \rho)$  &   \SI{5.5}{\tesla\meter}   \\
        Ion &   Au$^{65+}$  \\
         Energy $E$  &   \SI{150}{\mega\electronvolt\per\atomicmassunit}  \\
         Momentum Spread $\sigma_\delta$  &   \num{4.8e-4} \\
         Transverse emittances $\epsilon_{x,\text{norm}}^\text{4-rms}$, $\epsilon_{y,\text{norm}}^\text{4-rms}$  &  \SI{20}{\micro\meter}, \SI{4}{\micro\meter} \\
         \hline
         \hline
    \end{tabular*}
    \label{tab:BeamParameters}
\end{table}

\subsection{Measurement Uncertainty of Trajectory}
\label{Section:BPM_Resolution}
The trajectory of a single bunch can be recovered at the location of twelve beam position monitors (BPMs), which yield the horizontal and vertical position of the bunch centroid on a turn-by-turn basis. The position of single bunches at subsequent BPMs can then be identified and causally connected in order to construct a trajectory for each bunch across all sectors and over several turns after an oscillation is excited by an instantaneous kick. This is possible thanks to a sub-\si{\nano\second} synchronisation and the high resolution of the BPMs, which sample the beam position at \SI{250}{\mega\hertz}. %At speed of light, the passage time through one sector is \SI{60}{\nano\second}. This allows reconstruction of each of the four bunches centroid position sector-by-sector after their oscillations are excited by an instantaneous kick.

\begin{figure}
    \centering
    \includegraphics[width=\columnwidth]{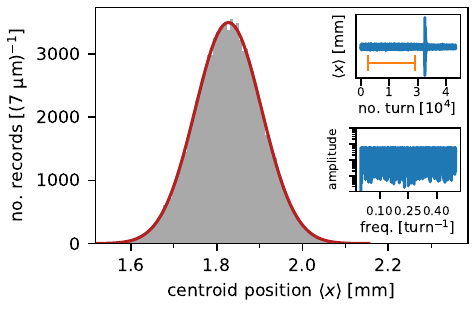}
    \caption{Single shot beam position monitor readings collected without beam excitation. The histogram of the measured centroid position follows a normal distribution, with its Gaussian fit displayed in red. The inset plot on the upper right displays the turn-by-turn data. The part of the cycle used for noise analysis, highlighted by the orange bar, comprises 100,000 data points collected from 4 bunches over 25,000 turns. The inset plot on the lower right displays a frequency spectrum of the turn-by-turn data utilized for noise estimation.}
    \label{Fig:BPM_PreNoise_Procedure}
\end{figure}

The BPMs installed at SIS18 are of shoebox type. 
Their measurement uncertainty is expected to mainly originate from thermal noise on the electrodes and thus, follow a normal distribution. The signal-to-noise ratio is affected by the beam current and amplifier settings. We estimate the uncertainty of the measured centroid position from measurement data. After sufficiently long time after the energy ramp, the beam dilutes into its equilibrium state before centroid oscillations are excited by the kicker magnet. The fluctuations of the recorded centroid positions follow a normal distribution with uniform frequency distribution, and we take its standard deviation as BPM resolution, cf.~Fig.~\ref{Fig:BPM_PreNoise_Procedure}.

The estimated measurement uncertainty of the bunch centroid in dependence of beam intensity is discussed in detail in ~\ref{App:BPM}. The majority of shots are recorded with a centroid position resolution of $\approx$\SI{150}{\micro\meter} in the horizontal plane, and \SI{75}{\micro\meter} in the vertical plane. 

\begin{figure}
    % \centering
    \includegraphics[width=\columnwidth]{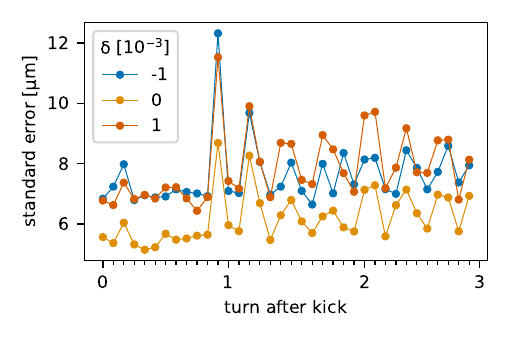}
    \caption{Standard error of the horizontal centroid position after centroid oscillations are excited by a transverse kick. The different colors represent an relative change in beam momentum $\delta$ originating from a deliberate mismatch of the rf-frequency w.r.t.\ to the revolution frequency.}
    \label{Fig:TrajStdErr}
\end{figure}

In order to achieve an improved resolution of the centroid trajectory, we average over different shots of the synchrotron. The uncertainty of the mean trajectory cannot be inferred directly from the BPM resolution since no analytic expression for the beam intensity distribution among many cycles is at hand. Instead, we observe that the recorded centroid position closely follows a normal distribution characterized by a mean $\mu$ and a standard deviation $\sigma$. Testing this hypothesis with a Shapiro-Wilk test \cite{Shapiro:ShapiroWilkTest} underpins that the data statistics does not deviate from a Gaussian normal distribution. Consequently, the sample mean $\hat{\mu}$ follows a normal distribution with standard deviation
\begin{align}
    \sigma_{\hat{\mu}} &= \frac{\sigma}{\sqrt{M}} \stackrel{M~\text{large}}{\approx} \frac{\hat{\sigma}}{\sqrt{M}} =: \hat{\sigma}_{\hat{\mu}}
\end{align}
where $M$ denotes the sample size. The standard error $\hat{\sigma}_{\hat{\mu}}$ estimates the uncertainty of the sample mean by replacing the unknown standard deviation of the distribution $\sigma$ by the standard deviation of the sample $\hat{\sigma}$. We use the standard error as measurement uncertainty of the averaged centroid trajectory used to reconstruct field errors, c.f.~Fig.~\ref{Fig:TrajStdErr}.

The standard error of the averaged trajectories increases in case the beam is radially displaced by mismatching the rf-frequency w.r.t.\ to the magnetic rigidity at flattop. We attribute this to beam losses caused by the beam halo touching the vacuum chamber. As a consequence, BPM noise is increased due to charged particles hitting its electrodes. In order to push the BPM performance to its limits a trade-off is required between high beam current yielding a high signal-to-noise ratio and a small transverse size enabling an excitation of the beam without losses. In the particular case of SIS18, the transverse beam size and injected beam current are proportional due to the used multi-turn injection scheme \cite{Appel:MTI}.
Furthermore, the standard error of the averaged trajectories increases after the beam completes one revolution. This is caused by the non-rectangular overshoot of the kicker pulse, see Appendix~\ref{App:QKicker} and c.f.\ Fig.~\ref{Fig:QKickerPulse}.
% {App:QKicker}

\subsection{Scenarios Investigated at SIS18}
The machine experiment is performed using the nominal doublet-likeoptics at extraction energy. Chromaticity is corrected by the foreseen two-family correction scheme featuring the two sextupole families \textsf{KS1C} and \textsf{KS3C}. 
We distinguish between the following two experimental scenarios:
\begin{enumerate}[A.]
    \item Chromaticity is corrected using all twelve lattice sextupoles.
    \item A single sextupole \textsf{GS09KS1C} is turned off while all other sextupoles are powered at the same currents as in (A.). This mimics a sextupole field error to be identified by the DLMN.
\end{enumerate}
Data for both scenarios are recorded \SI{36}{\hour} apart and analysed as two independent datasets, allowing to check how reproducible the DLMN training turns out w.r.t.\ to linear optics. The recorded datasets mainly differ in the intensity of the Au$^{65+}$ ion source, which was re-calibrated in between.

\subsection{Changes to Training Procedure}
In order to apply the DLMN method to SIS18, a few improvements are implemented compared to the original proposal of the method in \cite{Caliari:DLMN}.

\begin{enumerate}
    \item The SIS18 features a kicker magnet rotated by \SI{45}{\degree} allowing simultaneous beam excitation in both planes. Unfortunately, neither the exact rotation angle nor the magnetic field strength of the kicker are known precisely. Instead, we use the DLMN to back-propagate the centroid position measured at the first BPM downstream of the kicker location. A phase-space vector serving as initial condition, which is necessary to predict a trajectory with the DLMN, is inferred from 
    \begin{align}
        \vec{z} = (\text{DLMN})^{-1} \vec{z}_\text{BPM} \approx M^{-1} \vec{z}_\text{BPM} + \mathcal{O} \left( (\vec{z}_\text{BPM})^2 \right)
    \end{align}
    in first order approximation. Assuming zero initial transverse displacement, i.e.\ $\vec{z} = [0,p_x,0,p_y,0,\delta]^T$, the effective kick angles are calculated.    
    The first-order approximation $M$ of the DLMN is equivalent to the linear transfer matrix from the location of the kicker magnet to the first downstream BPM.
    
    The obtained transverse momenta $p_x$, $p_y$ of the initial condition depend on DLMN model parameters as there a several magnets in between kicker and BPM. Thus, the initial conditions are updated during each iteration of the training procedure.
    \item The beam momentum offset $\delta$ is varied through an offset of the rf-frequency. Both quantities are linked via the relation
    \begin{align}
        \frac{\Delta f_\text{rf}}{f_\text{rf}} = \eta \delta = \left( \alpha_c - \frac{1}{\gamma^2} \right) \delta 
    \end{align}
    by the momentum compaction factor $\alpha_c$ of the accelerator and the Lorentz-factor $\gamma$ of the beam. $\alpha_c$ describes the change in path length originating from transverse dispersion. In contrast to the original simulation study, we update our momentum offset $\delta$ based on the set rf-frequency and the $\alpha_c$ predicted by the DLMN during each iteration of the training procedure. This is necessary because the fractional momentum deviation is not a tuneable parameter, but rather a consequence of the set rf-frequency and the implicitly varying momentum compaction factor.
    \item We subtract the average centroid position from the time series at each BPM, as obtained from the noise analysis, c.f.\ Sec.~\ref{Section:BPM_Resolution}. This serves to keep the final optics model free from the influence due to (i) BPM misalignments, (ii) errors in the zero-reference calibration and (iii) shot-to-shot deviation of the closed orbit. This step requires subtraction of the transverse dispersion offset as predicted by the DLMN at the respective iteration of the training procedure.
\end{enumerate}

\section{DLMN Optics obtained by Training}
\label{Section:Results}
\label{Section:OpticsResults}
% \begin{figure}
%     \centering
%     \includegraphics[width=\columnwidth]{MAE_evol.pdf}
%     \caption{Evolution of mean absolute error (MAE) between measured trajectories and DLMN predictions. The gray line denotes the MAE between the standard error of the measured trajectories and zero.}
%     \label{Fig:MAE_Evol}
% \end{figure}

The quadrupole degrees of freedom of the DLMN comprise the normal and skew quadrupole components of the main lattice quadrupoles, i.e.\ the \textsf{QS1F}, \textsf{QS2D} and \textsf{QS3T} magnets per cell. The DLMN training procedure varies 72 quadrupole components in order to reproduce the measured trajectories. 

During training of the DLMN we observe convergence of the predicted to the measured trajectories by means of the mean-absolute error, e.g.\ Eq.~\eqref{Eqn:MAE}.
Because of the short time scale of seven turns and measurement uncertainties of the trajectory, c.f.\ Sec.~\ref{Section:BPM_Resolution}, the results may be subject to overfitting. The obtained DLMN optics therefore requires validation.

The training data set collected consists of 9 trajectories in total. 
They are obtained from five different excitation strength settings for the kicker magnet. 
Additionally, for the 3rd and 1st largest kick strengths, trajectories are recorded in case the beam momentum is altered by \num{\pm 1e-3}.

\subsection{Linear Optics}
\begin{figure}
    \centering
    \includegraphics[width=\columnwidth]{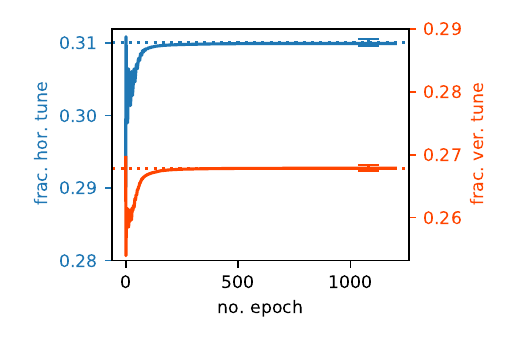}
    \caption{Evolution of horizontal (blue) and vertical (orange) tune during training of the DLMN. The dotted lines display the measured tunes and the errorbars indicate their uncertainty.}
    \label{Fig:TuneEvol}
\end{figure}

The recorded trajectories enable a tune measurement and allow to infer phase advances per sector. We apply the NAFF algorithm \cite{Laskar:NAFF} to the reconstructed sector-by-sector trajectory in order to measure the betatron tune, which is capable of achieving a higher frequency resolution than an FFT. The sample length is limited to \num{2048} turns after excitation by the kick because of signal decoherence. The tunes predicted by the DLMN converge against the measured tunes within their measurement uncertainty after few iterations, c.f.~Fig.~\ref{Fig:TuneEvol}.

The phase advance per sector is calculated by comparing the betatron oscillations phase at two adjacent BPMs. The phase is obtained from the complex Fourier coefficient
\begin{align}
    a_z = \frac{1}{N} \sum_{n=1}^N z_n \cdot e^{-2\pi i Q_z n}
    \label{Eqn:FourierCoefficient}
\end{align}
related to the tune $Q_z$ determined by NAFF. Here, $z$ denotes either the horizontal $x$ or vertical $y$ plane.

\begin{figure}
    \centering
    \includegraphics[width=\columnwidth]{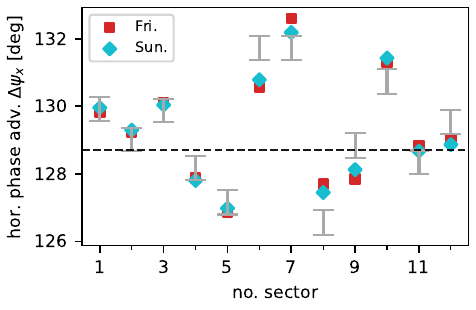}
    \caption{Horizontal phase advance measured from turn-by-turn BPM data (gray) compared to DLMN predictions. The training results are obtained from two independent measurements performed on Friday (red) and Sunday (cyan). The black dashed line displays the the phase advance predicted by the nominal accelerator model.}
    \label{Fig:PhaseAdvance}
\end{figure}

A comparison of observed and DLMN predicted phase advances shows reasonable agreement, c.f.~Fig.~\ref{Fig:PhaseAdvance}. Two independent training sets for the DLMN were collected on Friday afternoon and Sunday morning. 
The SIS18 was operated based on the same settings during both shifts, whilst on the Sunday shift the Au65+ source yielded a higher beam current. The difference in phase advance estimated from Fourier coefficients is negligible. The phase advances recovered from DLMN training match each other. The largest discrepancies are observed immediately downstream of the kicker magnet which is installed in sector 5.

\begin{figure}
    \centering
    \includegraphics[width=0.99\columnwidth]{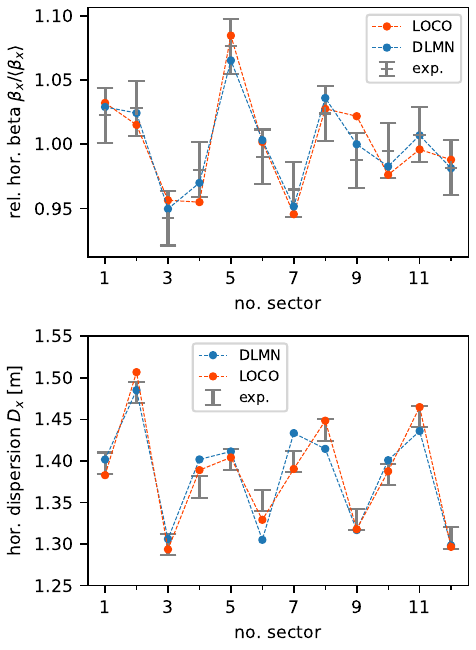}
    \caption{Horizontal beta-function (top panel) and dispersion (bottom panel) evaluated at BPM locations from spectral analysis of turn-by-turn data (gray). For comparison, the DLMN predictions (blue) and LOCO results (orange) are displayed by dots. }
    \label{Fig:Abs_BetxDx}
\end{figure}

The beta-function can be estimated from the magnitude of Eq.~\eqref{Eqn:FourierCoefficient} but suffers a systematic reduction for all BPMs due to decoherence.
We, therefore, compare beta-functions normalized to their mean, which cancels out the (averaged) centroid action.
Error bars take into account a variation in spectral amplitude originating from spectral noise.
In order to compare the relative course of the horizontal beta-function to absolute values predicted by either a trained DLMN or a LOCO fit of the ORM, we multiply by the average beta-function as predicted by the DLMN.
The horizontal dispersion is obtained from the closed orbit variation induced by the change in beam energy by mismatching the rf-frequency. The measurement uncertainty is deduced from the shot-by-shot rms closed orbit fluctuation, while the impact of turn-by-turn BPM noise becomes negligibly small due to the \SI{25}{\milli\second} long integration interval.
In the comparison between the spectral amplitude analysis of trajectory data and the DLMN predictions, we find an excellent agreement of beta-functions and dispersion, in particular in view of the relative differences between BPMs, c.f. Fig.~\ref{Fig:Abs_BetxDx}.

The orbit response matrix (ORM) is a criterion to verify whether the DLMN predictions interpolate well between the $i$th beam position monitor location where the centroid motion is observed. Each ORM matrix element
\begin{align}
    \text{ORM}_{ij} = \frac{\partial z_i}{\partial \Theta_j} = \frac{\sqrt{\beta^{(i)} \beta^{(j)}}}{\sin 2\pi Q} \cos \left( \pi Q - \left| \psi^{(i)} - \psi^{(j)} \right| \right)
\end{align}
incorporates beta-function and phase advance at the location of the $j$th steerer magnet. 
An ORM was measured in a subsequent beam time \cite{Isensee:ORM}. Here, $z$ denotes the change in either horizontal or vertical orbit caused by the deflection $\Theta_j$.
Unfortunately, no statistical evaluation of the orbit response w.r.t.\ to steerer magnet settings was performed.
Instead we estimate the measurement uncertainty of the ORM matrix elements from the observed shot-by-shot orbit fluctuations of about $\approx$\SI{100}{\micro\meter} to amount to \SI{0.1}{\meter\per\radian}.
The discrepancy between measured ORM and the DLMN prediction appears to be normal distributed with standard deviation $\sigma \approx$\SI{4e-2}{\meter\per\radian}. Hence, the ORM obtained from the DLMN corresponds to the measurement within its uncertainty.

The LOCO approach identifies a linear optics model by fitting a measured ORM. In the case of SIS18, we vary the individual strength of the \num{36} upright lattice quadrupoles. The beta-functions and dispersion predicted by LOCO agree well with the DLMN prediction, c.f.\ Fig.~\ref{Fig:Abs_BetxDx}.

We conclude that the DLMN prediction is not only meaningful at the location of beam position monitors, but also provides a good reconstruction of beta-functions and phase advances at the location of the steerer magnets.
The DLMN prediction may be harnessed for closed orbit control.

We observe that the identified DLMN optics coincide well with measured betatron tunes, phase advances, beta-functions and dispersion.
This demonstrates that the DLMN method is capable of reconstructing a valid linear optics model from short-term trajectory data evaluated over seven turns.
The DLMN's prediction for the linear optics model is of similar quality in case only a single trajectory is used for training. 
No scan of either the beam momentum or the kicker magnet's excitation strength are necessary, which underpins the potential saving in beam time.

\subsection{Non-linear Optics originating from Sextupoles}
The SIS18 features 12 lattice sextupoles located in each odd-numbered sector, which are available for chromaticity correction. The typically applied chromaticity correction scheme in the SIS18 control system consists of two families with six sextupoles each. We power the sextupoles accordingly, and then turn off a single sextupole in order to break the symmetry of the ring on purpose. This scenario serves as a test bed to demonstrate that the DLMN
\begin{enumerate}
    \item recovers the correct chromaticity as a global quantity
    \item localizes non-linear magnetic field errors. 
\end{enumerate}

The influence of non-linearities like sextupole magnets on the centroid trajectory can be estimated by normal form analysis. Ref.~\cite{Bartolini:NFA_BeamData} applies the non-resonant normal form to express the beam position in presence of non-linearities as
\begin{align}
    x(T) = \sqrt{\beta_x}& \mathrm{Re} \left[ \sqrt{2I_x} e^{i (2\pi Q_x T + \psi_x)} \right. \notag\\
    &- 2i \sum_{jklm} j f_{jklm} \left(2I_x\right)^{\frac{j + k - 1}{2}} \left(2I_y\right)^{\frac{l + m}{2}} \notag\\
    &\times \left. e^{i (1 -j+k) (2\pi Q_x T + \psi_x) } e^{i (l - m) (2\pi Q_y T + \psi_y) } \right] \quad ,
    \label{Eq:SpectralPosition}
\end{align}
far from resonant tunes. Eqn.~\ref{Eq:SpectralPosition} expresses the horizontal position $x$ at turn $T$ in terms of the (non-linear) actions $I_x$, $I_y$ and initial phases $\psi_x$, $\psi_y$. The amplitudes $f_{jklm}$ of harmonics of the betatron oscillation are referred to as \emph{resonance driving terms}. Each driving term is linked to a resonance condition in the Hamiltonian governing the equation of motion, and their magnitude depends on the set working point. The $f_{3000}$ represents the driving strength of the $3 Q_x = 13$ resonance, which is the closest third-order resonance to the presently configured working point at ($Q_x=4.31$, $Q_y=3.27$) including the identified discrepancies to the control system settings. Thus, $f_{3000}$ is the most relevant driving term to this experiment.

%OBF: Was genau ist eine 2Q_x Oszillation ? 
According to the nominal SIS18 optics model, the resulting amplitude of the $2Q_x$ oscillation driven by $f_{3000}$ is between \SI{3}{\micro\meter}-\SI{7}{\micro\meter} for the maximum available kick strength.
Hence, the non-linear motion is a small quantity compared to the measurement uncertainty of the beam position monitors $\sigma \approx$\SI{150}{\micro\meter}.
This underpins the necessity to suppress BPM noise by averaging trajectory data over many shots, resulting in a standard error roughly the size of the quantity of interest, c.f. Fig.~\ref{Fig:TrajStdErr}.

\begin{figure}
    \centering
    \includegraphics[width=\columnwidth]{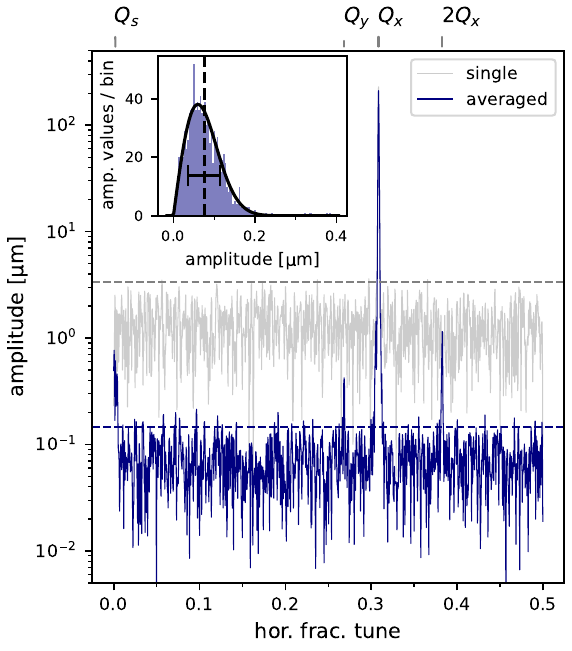}
    \caption{The horizontal centroid position's spectra are displayed in case a single synchrotron shot (gray) vs. an average over \num{506} shots (blue) is analyzed. A Hanning window is applied to suppress spectral leakage of the dominant peak originating from the horizontal betatron oscillation. The spectral noise is subject to a Rayleigh distribution displayed in the inlet plot. A fit of the distribution yields its expectation value and standard deviation denoted by the vertical resp. horizontal line. The dashed lines denote the distribution's 95th percentile characterizing the noise floor. After averaging, the horizontal betatron oscillation's second harmonic at $2Q_x$, the synchrotron oscillation at $Q_s$ and the vertical betatron oscillation at $Q_y$ are distinguishable from noise.}
    \label{Fig:TrajSpectrum}
\end{figure}

The resonance driving terms can be determined from spectral analysis of BPM turn-by-turn data according to Eq.~\eqref{Eq:SpectralPosition}. This requires that the amplitude of the corresponding spectral line be distinguishable from the noise floor which is distributed according to a Rayleigh distribution, c.f. Eq.~\eqref{Eqn:RDTNoiseFloorDistribution}.
The noise level decreases $\propto 1/ \sqrt{2N}$ with increasing sample length $N$. In the case of SIS18, the sample length is limited to \num{2048} turns because of decoherence. We characterize the noise floor by the 95th percentile of the noise distribution. In case of the trajectory shown in Fig.~\ref{Fig:TrajSpectrum}, averaging over $\approx$\num{500} shots reduces the noise floor from \SI{3.4}{\micro\meter} to \SI{0.14}{\micro\meter}. 

\begin{figure}
    \centering
    \includegraphics[width=\columnwidth]{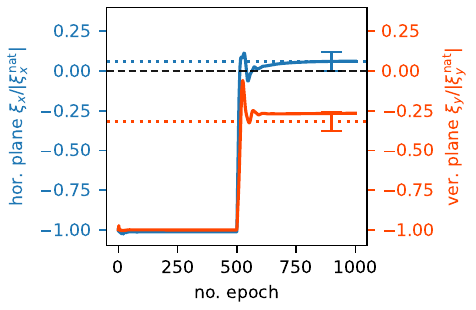}
    \caption{Horizontal (blue) and vertical (orange) chromaticity normalized to the natural chromaticity as calculated from the nominal optics model. The measured normalized chromaticities are marked by the dotted lines, solid lines display the DLMN prediction during training. Chromaticity is corrected according to the nominal optics model setting, shown by the black dashed line. The DLMN varies each of the twelve lattice sextupoles individually.}
    \label{Fig:ChromaEvol}
\end{figure}

The DLMN is initialized at natural chromaticity without any prior knowledge about the non-linear magnetic fields. In addition to the normal and skew quadrupole strength of three quadrupoles per lattice cell, we allow the DLMN training procedure to vary the sextupole strength of each lattice sextupole individually. With twelve sextupole degrees of freedom, the DLMN predicts a horizontal chromaticity that agrees well with the measured one ($\Delta \xi_x / \xi_x^\text{nat} =$\SI{0.6}{\percent}), c.f.~Fig.~\ref{Fig:ChromaEvol}.
The predicted vertical chromaticity is slightly exceeds the measurement ($\Delta \xi_y / \xi_y^\text{nat} =$\SI{5}{\percent}), but is more accurate compared to the nominal accelerator model $(\Delta\xi_x / \xi_x^\text{nat}, \Delta\xi_y / \xi_y^\text{nat}) =$ (\SI{6}{\percent},\SI{33}{\percent}).

% The large number of sextupole degree of freedom may result in overfitting issues, as the DLMN training procedure uses these localized magnetic field strengths in order to fit measurement errors of the trajectory. Considering a DLMN featuring only two degrees of freedom for the two families of sextupoles (with constant strength for all sextupoles within a family), the variation of sextupole strengths directly affects the predicted trajectories globally. However, the predicted chromaticities also differ slightly from the measured ones, c.f.~Fig.~\ref{Fig:ChromaEvol_TwoFamilies} although the agreement is better than for 12 individual sextupole degrees of freedom. The difference in predicted horizontal chromaticity is negligible, whereas the predicted vertical chromaticity is closer to its measured value in case of less degrees of freedom.

% \begin{figure}
%     \centering
%     \includegraphics[width=\columnwidth]{Figure_8.pdf}
%     \caption{Horizontal (blue) and vertical (orange) chromaticity normalized to the natural chromaticity calculated from the nominal optics model. The measured normalized chromaticities are marked by the dotted lines, solid lines display the DLMN prediction during training. Chromaticity is corrected according to the nominal optics model, shown by the black dashed line. The sextupoles are grouped into two families and hence, the DLMN varies only two sextupole degrees of freedom.}
%     \label{Fig:ChromaEvol_TwoFamilies}
% \end{figure}

Decoherence decreases the spectral lines amplitude, which introduces a systematic underestimation of the resonance driving terms.
This systematic error may be corrected if the decoherence of the centroid oscillation can be quantified. Instead, we evaluate the relative resonance driving terms w.r.t.\ to their mean, which is independent of the centroid oscillations mean amplitude over the FFT window.
The relative strength of the local resonance driving term $f_{3000}$ indicates the location of sextupole errors up to the section between two adjacent BPMs.

The value of $f_{3000}$ is determined at the location of BPMs for the case of the regular chromaticity correction scheme where two families of sextupoles are employed, except a single sextupole \textsf{GS09KS1C} being turned off in sector 9.
The predicted effect on $f_{3000}$ is a local increase at the location of the perturbation in Sector 9 and a consecutive decrease on the opposite site of the ring in Sector 3.

\begin{figure}
    \centering
    \includegraphics[width=\columnwidth]{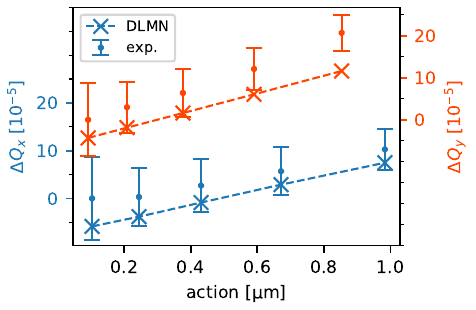}
    \caption{Comparison between simulated and measured (errorbars) resonance driving terms $f_{3000}$. The black line corresponds to the nominal accelerator model based on MAD-X. The red line displays the DLMN prediction after training. The symmetry breaking is caused by a single sextupole \textsf{GS09KS1C} being turned off in sector 9.}
    \label{Fig:f3000_Comparison}
\end{figure}

The measured resonance driving term $f_{3000}$ is compared to predictions from the nominal accelerator model and the DLMN after training, c.f.~Fig.~\ref{Fig:f3000_Comparison}.
The perturbation to $f_{3000}$ in sector 9 and its mirroring on the opposite site in sector 3 agree well with the nominal accelerator optics. The DLMN predicts a large increase in $f_{3000}$ in sector 9 which agrees qualitatively with the measurement.
A quantitative difference however is larger than our uncertainty estimate of the measured resonance driving terms in sectors 10 and 12.
This is most likely related to the DLMN training set comprising only seven turns compared to 2048 turns used as FFT sample length, as stochastic measurement errors do not average out over few data points.
The DLMN training procedure targets at a minimal discrepancy between predicted and measured trajectories, and thus DLMN degrees of freedom are adjusted to reproduce erroneous beam positions.

\begin{figure}
    \centering
    \includegraphics[width=\columnwidth]{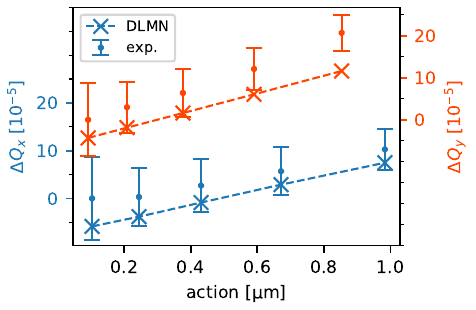}
    \caption{Comparison of the horizontal (blue) and vertical (orange) amplitude detuning between measurement and DLMN prediction. The tunes are displayed w.r.t.\ to the action of the beam centroid induced by the kicker magnet, and the tune of the lowest excitation is substracted.}
    \label{Fig:AmpDetuning}
\end{figure}

Moreover, the amplitude detuning is an observable to judge the quality of the trained DLMN. 
Trajectories recorded at five different excitation strengths of the kicker magnet in combination with precise algorithms to determine the betatron tune \cite{Zisopoulos:MixedBPM_NAFF} reveal an increase of the tunes with the beam centroid's action in both planes. 
The tunes predicted by the DLMN show a similar amplitude detuning, with excellent agreement in the horizontal plane, c.f.\ Fig.~\ref{Fig:AmpDetuning}. 

The training results presented can be reproduced by training on \num{3} trajectories recorded for similar excitation strengths with similar quality.
Application of the DLMN thus only requires a scan of the fractional beam momentum offset $\delta \in [-10^{-3}, 0, 10^{-3}]$.

\subsection{Possible Applications of Trained DLMN: Tunescans}
One key advantage of the DLMN method stems from its physically meaningful degrees of freedom, which are magnetic multipole components.
This warrants further use of the training results to perform either analytic or simulation studies.
The obtained refined optics model can be converted to lattice files of common tracking codes like MAD-X \cite{Berg:MAD-X} or XSuite \cite{Iadarola:XSuite}. 

Using the case of SIS18 as an example, we measure the qualitative effect of resonances with a dynamic tune scan. A beam comprising \num{1.2e10} protons is accelerated from \SI{11.4}{\mega\electronvolt\per\atomicmassunit} to \SI{400}{\mega\electronvolt\per\atomicmassunit} to suppress the influence of space charge present at injection energies. The transverse emittance before blow up at the vertical integer resonance and momentum spread are $\epsilon_{x,\text{geo}}^{1-\text{rms}} \approx$\SI{1}{\micro\meter} resp.\ $\Delta p/p \approx$\num{3e-4}. All lattice sextupoles are turned off, and observed resonances are attributable to unknown magnetic field errors.
At constant energy, the continuously stored bunched beam is first increased in transverse size by moving the working point close to the vertical integer resonance $3Q_y=9$ causing the beam emittance to grow. Just when first beam loss is encountered, the working point is moved vertically upward towards the vertical half-integer resonance. Each resonance crossed in this procedure causes a peak in normalized loss rate $\frac{1}{I} \left| \frac{\partial I}{\partial t} \right|$. The losses are recorded by measuring the instantaneously stored beam current $I$ by means of a DC current transformer. This approach is repeated for many different horizontal tune settings to cover a 2D area in the transverse betatron tune space. 

\begin{figure*}
    \centering
    \includegraphics[width=\textwidth]{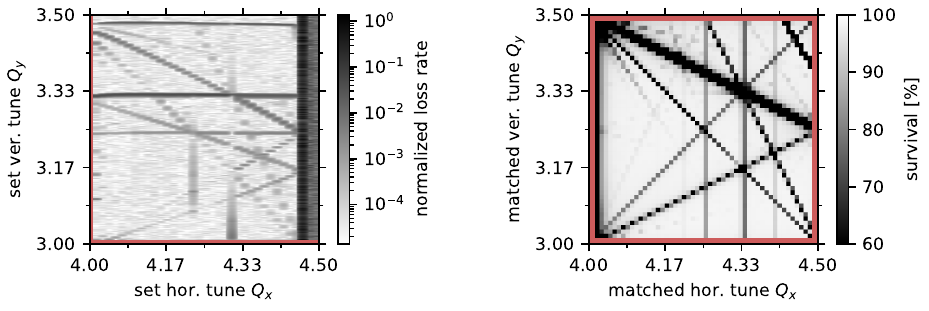}
    \caption{Comparison of measured dynamic tune scan (left) and simulated static tune scan based on DLMN predictions (right). Set tune refers to the tune specified in the accelerator control system. The systematic tune shift between nominal accelerator optics model and measurements is evident from the location of sextupole resonances. In case of the simulated tune scan based on DLMN optics, matched tune refers to the working point passed to the XSuite matching routine. }
    \label{Fig:TuneScan_Comparison}
\end{figure*}

The results of this dynamic tune scan for SIS18 are presented in Fig.~\ref{Fig:TuneScan_Comparison} in the left panel. 
They confirm the shift in tune $\Delta Q_x\approx \num{2e-2}$, $\Delta Q_y\approx \num{1e-2}$ between set and actual value.
Consequently, the normal sextupole resonance (3,0,13) is found below rather than above $Q_x=4.33$, c.f.\ Fig.~\ref{Fig:TuneScan_Comparison} in the left panel.
Here, we use the notation ($a$,$b$,$c$) to indicate resonance condition $a Q_x + b Q_y = c$ fulfilled by integers $a,b,c \in \mathbb{Z}$.
Furthermore, the measurements reveal the presence of excited skew sextupole resonances. They may originate from non-zero linear coupling.

A previous tune scan performed at SIS18 at injection energy is reported in Ref.~\cite{Franchetti:Tunescan}, where space charge effects were suppressed by using low intensities. The scan is subject to various systematic betatron tune shifts w.r.t.\ to the now deprecated previous control system named SISMODI. In agreement with our dynamic tune scans, the presence of linear coupling resonance as well as 3rd-order resonances originating from upright sextupoles is observed. Furthermore, the skew sextupole and octupole resonances at (0,3,10) resp.\ (0,4,13) are detected by both scans. 

A simulation study to investigate the origin of these resonances and possible mitigation schemes may be performed by means of DLMN training results.
The DLMN predictions for normal and skew quadrupoles, as well as normal sextupoles are loaded into XSuite, and a simulated static tune scan is performed.
The simulation tracks a bunched beam at natural chromaticity.
The static tunescan is very sensitive to the difference resonance (1,-1,-1) because of the large emittance of the used U$^{73}$ beam foreseen for the FAIR project. All experimentally observed normal sextupole resonances are excited, with the $(1,2,11)$ resonance being stronger than the $(1,-2,-2)$ resonance in agreement with the measurement. The skew sextupole resonance $(2,1,12)$ is excited by normal sextupole errors in combination with skew quadrupole errors which provide linear coupling. This is verified by comparison to a resonance diagram computed with all DLMN predicted skew quadrupole components being removed.
Hence, it is plausible that their excitation can be cured by correcting linear coupling and restoring symmetry of the $f_{3000}$ resonance driving term.
Several higher order resonances observed in the measured dynamic tune scan are excited by the DLMN quadrupole and sextupole field error predictions.
Most notably, the magnetic field errors recovered by the DLMN drive the normal octupole resonances (4,0,17) and (2,2,15), which are clearly excited in the measurements.
A major difference to the DLMN error model is the observation of skew sextupole and skew octupole resonances (0,3,10) and (0,4,13), indicating the presence of a source of skew sextupole errors in SIS18.
Related peaks in the spectrum of BPM turn-by-turn data were not detected and are likely covered by the noise floor. 
One possibility to hunt skew sextupole errors by the DLMN approach is to move the working point closer to the related resonances and thus, increase the magnitude of their driving terms.

The study of resonances by means of comparing simulated tune scans to measurements is exemplary for possible applications of DLMN training results.  The non-linear optics model in terms of magnetic multipole components predicted by the DLMN can readily be used for simulation studies. The DLMN results may be applied to identify the origin of resonances and enable their compensation.

\section{Conclusion}
\label{Section:Summary}
The DLMN approach to identify magnetic field errors has been applied in a hadron synchrotron experiment for the first time. A refined optics model based on the measured beam motion has been created for the SIS18 synchrotron. One possible application of the DLMN training results is presented in terms of resonance analysis by means of tracking simulations, which make use of the DLMN identified field error multipole values.

In contrast to established alternative approaches, the DLMN reconstructs both linear and non-linear field errors in parallel.
Its application does not require an accurate linear optics model in before.
The learned linear optics model is observed to accurately recover tunes, beta-functions and dispersion. Predicted phase advances agree well with a spectral amplitude analysis of beam position monitor turn-by-turn data.
The accelerator optics predicted by the DLMN are consistent with a LOCO-fit of a separately measured orbit response matrix (ORM). Unlike the ORM measurement, the DLMN does not require a systematic scan of steerer magnet deflection angles.
The construction of a linear optics model requires measurement of a single trajectory only.

The reconstruction of sextupole degrees of freedom is limited by the measurement uncertainty, which is in case of the considered synchrotron of similar magnitude as the non-linearities of interest.
The effective lattice model established via DLMN exhibits approximately correct chromaticities, that is a deviation of $(\Delta\xi_x / \xi_x^\text{nat}, \Delta\xi_y / \xi_y^\text{nat}) =$ (\SI{0.6}{\percent},\SI{5}{\percent}) vs.\ $(\Delta\xi_x / \xi_x^\text{nat}, \Delta\xi_y / \xi_y^\text{nat}) =$ (\SI{6}{\percent},\SI{33}{\percent}) of the control system.
Furthermore, the spatial evolution of resonance driving term $f_{3000}$ is qualitatively recovered. 
However, the length limitation of the utilized trajectories to seven turns makes the DLMN more prone to beam position monitor noise than FFT-based methods which typically take into account hundreds of turns.

The DLMN approach yields an effective accelerator model in terms of magnetic multipole components. This enables further study of the DLMN predicted refined optics model (and lattice) by means of tracking codes and similar established tools in accelerator physics.  The refined optics model enables a more detailed study of resonance induced beam loss with a more accurate accelerator behaviour. 
It may be applied to correction of linear coupling and resonance compensation.

The limited amount trajectories, one in case of linear and three in case of non-linear optics, required to train the DLMN allow its integration into regular operation. 
A monitoring of the accelerator's state and optics on a daily basis is thus possible.

\section*{Acknowledgements}
The authors thank N.\ Madysa for technical advice regarding data acquisition and O.\ Chorniy for extensive support on operating the beam position monitors and the SIS18 and UNILAC operating crews. We thank V. Isensee for the contributed orbit response matrix measurement.

\appendix

\section{Beam Position Monitor Resolution}
\label{App:BPM}

The measurement uncertainty of the centroid position is subject to systematic errors originating from misalignments and thermal noise. A lateral shift of the BPM electrodes is negligible to this measurement, as it only affects the measured equilibrium beam position. During this measurement, only the dynamic motion of the beam centroid around the equilibrium after excitation is of interest, and the closed orbit is disregarded.

A potential issue are rotations of the BPMs with respect to the laboratory reference system. Such misrotated BPMs would detect an erroneous coupling between the horizontal and vertical centroid motion. The BPMs are included in the components that are precisely aligned, and we neglect BPM misrotations in our analysis.

The stochastic measurement uncertainties are expected to primarily originate from thermal noise on the electrodes, which is expected to be normal distributed. Crucial to the resolution of the beam centroid is the signal-to-noise ratio. The noise contribution is expected to be proportional to temperature. The signal strength is proportional to the beam current and the centroid oscillation amplitude. In case the beam rests in its equilibrium state, the noise level can be estimated from the fluctuating beam position, c.f.~Fig.~\ref{Fig:BPM_PreNoise_Procedure}. The observed fluctuations are consistent with a normal distribution.

\begin{figure*}
    \centering
    \includegraphics[width=0.8\textwidth]{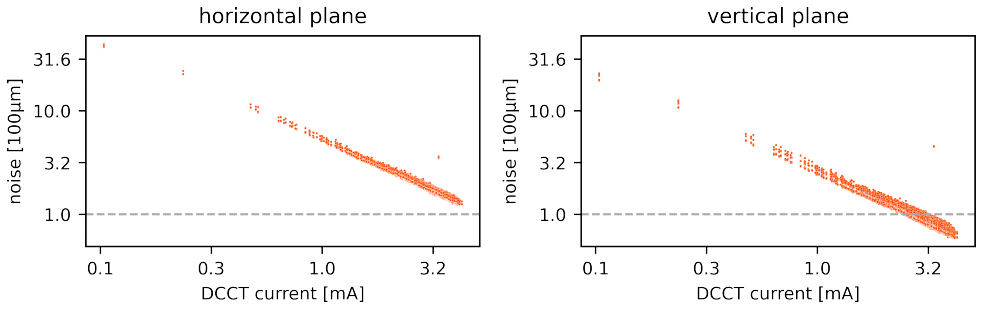}
    \caption{Standard deviation of the normal distributed noise of the centroid position vs. beam current. A large beam current suppresses fluctuations of the beam position reading while the beam is in its equilibrium state.}
    \label{Fig:Intensity_vs_Noise}
\end{figure*}

The uncertainty depends on the beam current resulting in a variation of the resolution performance of the BPMs on a shot-to-shot basis, c.f. Fig.~\ref{Fig:Intensity_vs_Noise}.
The beam current at magnetic flattop varies due to fluctuations of the ion source, which is particulary true for the utilized Au$^{65+}$ source. 

The distribution of estimated BPM resolution per sector and plane is shown in ~\ref{App:BPM}. The majority of shots are recorded with a centroid position resolution of \SI{120}{\micro\meter} to \SI{180}{\micro\meter} in the horizontal plane, and \SI{60}{\micro\meter} to \SI{90}{\micro\meter} in the vertical plane. Shots with an estimated resolution worse than \SI{630}{\micro\meter}, which originate from very low beam currents, are excluded from further data analysis.

The fluctuating beam current is subject to an unknown distribution, and thus, an analytic calculation of the effect on the trajectory uncertainty is not feasible.

Instead, we empirically observe that the recorded beam position sector-by-sector downstream of the initial kick is normal distributed, with some mean $\mu$ and standard deviation $\sigma$, consistent with the Shapiro-Wilk test. The averaged trajectory is given by the sample mean $\hat{\mu}$, which is itself normal distributed with standard deviation
\begin{align}
    \sigma_{\hat{\mu}} &= \frac{\sigma}{\sqrt{N}} \notag\\
    &\approx \frac{\hat{\sigma}}{\sqrt{N}} =: \hat{\sigma}_{\hat{\mu}}
\end{align}
where $N$ denotes the sample size. The standard error $\hat{\sigma}_{\hat{\mu}}$ estimates the uncertainty of the sample mean by replacing the unknown standard deviation of the distribution $\sigma$ by the standard deviation of the sample $\hat{\sigma}$.

\begin{figure*}
    \centering
    \includegraphics[width=0.8\textwidth]{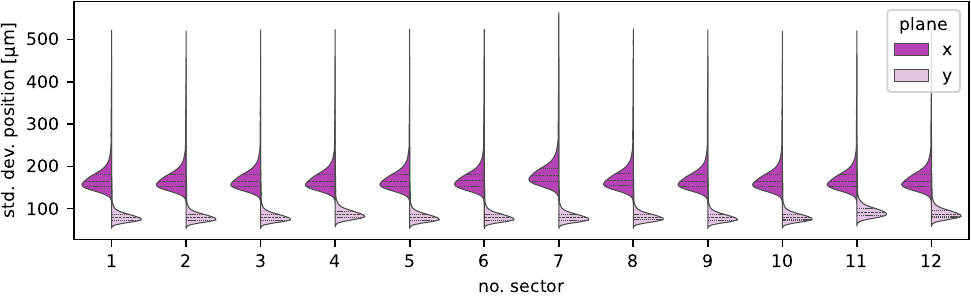}
    \caption{Estimated BPM resolution per sector and per plane (x/horizontal, y/vertical). Cycles where the estimated noise exceeds \SI{630}{\micro\meter} are excluded from this figure and any further data analysis.}
    \label{Fig:BPM_Noise_perSectorPlane}
\end{figure*}

The estimation of resonance driving terms requires a frequency analysis of BPM data and, hence, is subject to BPM noise. Assuming the BPM readings over $N$ turns are normal distributed
\begin{align}
    x_n \propto \mathcal{N}(0,\sigma)
\end{align}
the Fourier coefficients
\begin{align}
    a_k = \sum_{n=0}^{N-1} x_n e^{-2i\pi \frac{k}{N} n}
\end{align}
real and imaginary part are normal distributed as well
\begin{align}
    \mathrm{Re} (a_k), \mathrm{Im} (a_k) \propto \mathcal{N} \left( 0,\frac{\sigma}{\sqrt{2N}} \right) \quad .
\end{align}

Hence, the amplitude
\begin{align}
    |a_k| = \sqrt{\mathrm{Re}^2(a_k) + \mathrm{Im}^2(a_k)} \propto \text{Rayleigh}(\sigma)
\end{align}
is subject to a Rayleigh distribution. The expected noise amplitude is a systematic addition to the spectrum of centroid motion.

\section{Beam Excitation via Kicker Magnet}
\label{App:QKicker}

The dynamics of centroid motion are investigated by deflecting the beam out of its equilibrium state. This is done employing a fast kicker magnet which creates a magnetic field enduring roughly the revolution period of the particles. Ideally, each bunch is subject to a constant Lorentz-force during a single passage of the short kicker magnet, resulting in an instantaneous change of transverse momenta. This kick can be used as an initial condition for tracking. The ideal magnetic fields time dependence is therefore a rectangular pulse with zero time spent on the rising and falling flank. 

The SIS18 features a single kicker magnet rotated by \SI{45}{\degree} w.r.t. to the laboratory frame, located in sector 5. Because of the installation orientation both horizontal and vertical momentum are affected simultaneously. Furthermore, the power supply is unipolar, limiting the beam excitation to positive kick strengths.

\begin{figure*}
    \centering
    \includegraphics[width=0.8\textwidth]{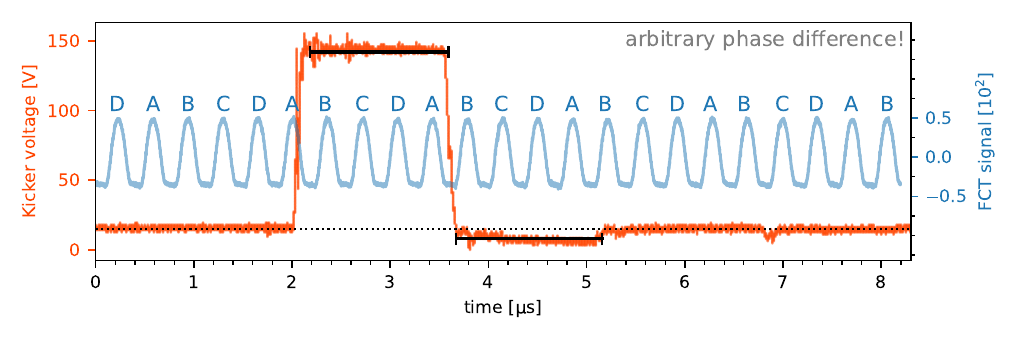}
    \caption{Time course of the kicker magnetic field (orange) compared to 
    longitudinal charge signal (blue) recorded by a fast current transformer (FCT). The black bars indicate the mean of the kicker voltage during the flattop resp.\ overshoot. The SIS18 hosts 4 bunches denoted by letters. The phase between kicker voltage and FCT signal is arbitrary since the oscilloscope recording the kicker signals runs independently of the accelerator timing system.}
    \label{Fig:QKickerPulse}
\end{figure*}

The pulse shape of the magnetic field can be inferred from a voltage signal proportional to the current, and recorded by an oscilloscope. Due to technical limitations, the oscilloscopes reading could only be stored as a screenshot, preventing a thorough statistical analysis of the pulse shape and estimation of noise among different cycles. The kicker pulse features steep rising and falling flanks, an approximately constant flattop and an overshoot, c.f.\ Fig.~\ref{Fig:QKickerPulse}.

\section{Determination of Resonance Driving Term $f_{3000}$}
The resonance driving term $f_{3000}$ is most relevant to this experiment as the SIS18 working point ($Q_x=4.31$, $Q_y=3.27$) is close to the associated $3Q_x = 13$ resonance.
Following Eq.~\ref{Eq:SpectralPosition}, its value is obtained by fitting the amplitude of the spectral line located at twice the betatron tune w.r.t.\ to the action of the beam centroid.

The amplitude of the spectral line $A$ is connected to the amplitude
\begin{align}
    \Tilde{A} = a c s A + A_\text{leak} + A_\text{noise} \; , \: a,c,s \in (0,1) \; ,
\end{align}
obtained from the Fourier coefficient of turn-by-turn BPM data.
Here, $a$ and $c$ denote a systematic reduction due to amplitude resp.\ chromatic detuning, $c$ the scalloping effect in case the Fourier coefficient is sampled at a slightly off frequency, $A_\text{leak}$ the spectral leakage from the dominant horizontal betatron ocsillation peak and $A_\text{noise}$ the contribution from BPM noise.

The scalloping effect is minimized by precise estimation of the betatron tune using NAFF, the leakage effect is suppressed by application of a Hanning window and both are neglect in further analysis.
One drawback of tackling spectral leakage using a window is its peak widening and reduction of amplitude, which further complicates estimating the absolute value of $A$.
The noise contribution is random, and its systematic effect canceled by subtracting the expectation value.
The standard deviation of the noise distribution is considered to be the measurement uncertainty. 

The systematic effects of amplitude and chromatic detuning cannot be estimated without knowledge of the beam distribution in phase space.
Instead, we evaluate relative resonance driving terms by dividing by the average amplitude across the synchrotrons circumference.
Because the decoherence time $t$ is larger than the revolution period $t \gg T_\text{rev}$, the factors $a$ and $c$ are the same at every location in the ring.

\begin{figure}
    \centering
    \includegraphics[width=\columnwidth]{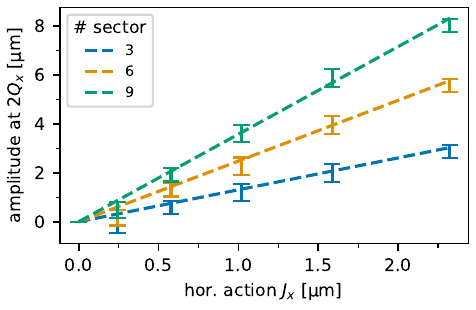}
    \caption{Dependence of the spectral line's amplitude at $2 Q_x$ on the horizontal action $J_x$ observed in different sectors of SIS18. The dependency is consistent with the expected linear increase in action, and the resonance driving term $f_{3000}$ obtained from the slope.}
    \label{Fig:F3000_AmpVsAction}
\end{figure}

The resultinng amplitudes increase linearly with the beam centroid's action, which agrees with the theoretical expression of Eqn.~\ref{Eq:SpectralPosition} derived from normal form analysis, c.f.\ Fig.~\ref{Fig:F3000_AmpVsAction}.
The estimation of relative resonance driving terms is independent from uncertainties regarding the action, but requires solid knowledge of the local beta function.

\bibliographystyle{elsarticle-num} 
\bibliography{Bibliography}



%% For example:

%% \bibitem[Aladro et al.(2015)]{Aladro15} Aladro, R., Martín, S., Riquelme, D., et al. 2015, \aas, 579, A101

%%\end{thebibliography}

\end{document}